\documentclass[aps,prd,preprint,groupedaddress,showpacs,showkeys]{revtex4}
\begin{document}
\title {Semiclassical (QFT) and Quantum (String) Rotating Black Holes and their Evaporation: New Results }

\author{A. Bouchareb$^{1,2}$, M. Ram\'on Medrano$^{3,2}$ and N.G. S\'anchez$^{2}$}
\email{Norma.Sanchez@obspm.fr,  mrm@fis.ucm.es}
 
\affiliation{
(1) Departement de Physique, Universit\'e d' Annaba, B.P.12, El-Hadjar, Annaba 2300, Algerie.\\
(2) Observatoire de Paris, LERMA, CNRS UMR 8112,  61, Avenue de l'Observatoire, 75014 Paris, France. \\
(3) Departamento de F\'{\i}sica Te\'orica, Facultad de Ciencias F\'{\i}sicas, Universidad Complutense, 
E-28040 Madrid, Spain}

\date{\today}
\begin{abstract}
Combination of both quantum field theory (QFT) and string theory in curved backgrounds in a consistent framework, the string analogue model, allows us to provide a full  picture of the Kerr-Newman black hole and its evaporation going {\bf beyond} the current picture. We compute the quantum emission cross section of strings by a Kerr-Newmann black hole (KNbh). It shows the black hole emission at the Hawking  temperature $T_{sem}$ in the early stage of evaporation and the new string emission featuring a Hagedorn transition into a string state of temperature $T_ {s}$ at the last stages. {\bf New} bounds on $J$ and $Q$ emerge in the quantum string regime (besides the known ones of the classical/semiclassical QFT regime). The last state of evaporation of a semiclassical KNbh with mass $M > m_{Pl}$, angular momentum  $J $ and charge $Q$ is a string state of temperature $T_{s}$, string mass $M_s$, $J = 0$ and $Q = 0$, decaying as usual quantum strings do into all kinds of particles.({\bf Naturally}, in this framework, there is {\bf no} loss of information, (there is {\bf no} paradox at all)). We compute the string entropy $S_s(m, j)$ from the microscopic string density of states of mass $m$ and spin mode $j$, $\rho (m,j)$. (Besides the Hagedorn transition at $T_{s}$), we find for high $j$, ({\bf extremal string} states $j\rightarrow m^2\alpha^{'}c $), a {\bf new} phase transition 
at a temperature $ T_{sj}~=~ \sqrt{j/\hbar}~T_s $, {\bf higher} than $T_{s}$. By precisely identifying the semiclassical and quantum (string) gravity regimes, we find a {\bf new} formula for the Kerr black hole entropy $S_{sem} (M,J)$, as a function of the usual Bekenstein-Hawking entropy $S_{sem}^{(0)}$. For $M\gg m_{mPl}$ and $J < GM^2/c$, $S_{sem}^{(0)}$ is the leading term, {\bf but} for high angular momentum, (nearly {\bf extremal} case $J = GM^2/c$), a gravitational phase transition operates and the whole entropy $S_{sem}$ is drastically  {\bf different} from the Bekenstein-Hawking entropy $S_{sem}^{(0)}$. This {\bf new} extremal black hole transition occurs at a temperature $ T_{sem J}~=~ \sqrt{J/\hbar}~T_{sem} $, {\bf higher} than the Hawking temperature $
T_{sem} $.
\end{abstract}

\maketitle

\section{Introduction and Results \label{sec:intro}}

The issue of the last stages of black hole evaporation requires a quantum gravity
description. In principle, this question should be properly addressed in the
context of String Field Theory, but there is no tractable framework for it (even in the simplest flat space situations without black holes). However, the string analogue model implemented in curved backgrounds, provides a suitable framework for this purpose. This approach allows to combine both quantum field theory (QFT) and string theory in curved backgrounds and to go further in the understanding of quantum gravity effects \cite{1}-\cite {4}. The string analogue model is particularly appropriated and natural for black holes, as Hawking radiation and strings posses intrinsic thermal features and temperatures.

\bigskip

In this framework, strings are considered as a collection of quantum fields $\Phi_n$ coupled to the curved background, 
and whose masses $m_n$ are given by the degenerate string mass spectrum in the curved space considered.
Each field $\Phi_n$ appears as many times the string degeneracy of the mass level $\rho (m)$. Although the
fields $\Phi_n$ do not interact among themselves, they do with the black hole background. The mass formula $m (n)$ and $\rho (m)$ are obtained by solving the quantum string dynamics in the curved background considered. The string density of levels $d(n)$ of level $n$ is the same in flat and curved space-time. The mass relation between $m$ and $n$, and thus $\rho (m)$, depend on the curvature of the background geometry and are in general different from the flat space-time string mass spectrum and flat space mass level density.

\bigskip

In this paper we go further in this research~\cite{1}- \cite{4}. We include the effects of the charge and angular momentum both for the black holes and strings. A clear picture for the Kerr-Newman black hole emerges, going {\bf beyond} the current picture, both for its semiclassical regime and for its quantum (string) regime.

\bigskip

We compute the quantum emission cross section of strings by a Kerr-Newmann black hole. For $T_{sem}\ll T_{s}$, that is,  in the early stage of evaporation, the string cross section shows the Hawking part of the emission with temperature $T_{sem}$, namely, the semiclassical or QFT regime. For $T_{sem}\rightarrow T_{s}$, the massive string modes dominate the emission, the string cross section shows a Hagedorn phase transition at  $T_{sem}$ = $T_{s}$. This phase transition undergone by the emitted strings represents the non-perturbative back reaction effect of the emission on the black hole. The last state of evaporation of a semiclassical Kerr-Newmann black hole with mass $M > m_{Pl}$, $T_{sem}(J,Q)~<~ T_{s}$, angular momentum  $J $ and charge $Q$ is a string state of string temperature $T_{s}$, string mass $M_s$, $J = 0$ and $Q = 0$, which decays by the usual quantum string decay into all kinds of particles. ({\bf Naturally}, in this framework,  there is {\bf no} loss of information, (there is {\bf no} paradox at all)). Besides the classical/semiclassical known bounds on $J$ and $Q$, {\bf new} bounds  emerge in the quantum string regime. 

\bigskip

A central object is  $\rho (m,j)$, the microscopic string density of states of mass $m$ and spin mode $j$.  We find for high $j$,  that is,  for the {\bf extremal string} states $j\rightarrow m^2\alpha^{'}$, a {\bf new} phase transition at a temperature $T_{sj} = \sqrt{(j/\hbar)}T_s$, higher than $T_{s}$. We call it {\bf extremal transition}. This is {\bf not} the usual Hagedorn transition, which occurs here as well for high $m$ at the temperature $T_{s}$. The characteristic behaviour of the extremal transition is a square root branch point near $T_{sj}$. It manifests, in particular, as a logarithmic singularity in the string entropy $S(m,j)$. This extremal behaviour is {\bf universal} (it holds in any number of dimensions) and is analogous to the transition found for the thermal self-gravitating gas of point particles  \cite{5} and for strings in de Sitter background \cite{4}. In our case here, $j\neq 0$, which acts in the sense of the string tension, appears in the transition as an "smaller  $\alpha^{'}$ ", ( $\alpha^{'}_j~  \equiv ~ \sqrt{\hbar/j}~ \alpha^{'}$"') ~ , and thus a {\it higher} tension.

\bigskip

By precisely identifying the semiclassical and quantum (string) gravity regimes, we find a {\bf new} formula for the Kerr black hole entropy $S_{sem} (M,J)$, which is a function of the usual Bekenstein-Hawking entropy $S_{sem}^{(0)}$.
For $M\gg m_{mPl}$ and $J < GM^2/c$, $S_{sem}^{(0)}$ is the leading term of this expression, {\bf but} for high angular momentum, (nearly {\bf extremal} or extremal case $J= GM^2/c$), a gravitational phase transition operates and the whole entropy $S_{sem}$ is drastically  {\bf different}  from the Bekenstein-Hawking entropy $S_{sem}^{(0)}$. This {\bf new phase transition} takes place at a temperature $T_{sem J}= \sqrt{(J/\hbar)}~T_{sem} $, {\bf higher} than the Hawking temperature $T_{sem}$.
The characteristic features of this transition are similar to those of the extremal string transition we found here for the extremal string states. Our discussion on the string case translates to the semiclassical black hole quantities but we do not extend more on it here. 

\bigskip

This work does not make use of AdS (Anti-de Sitter space), neither of CPP's (conjectures, proposals, principles, formulated in string theory in the last years).

\section{Kerr-Newman Black Hole\label{sec:KNBH}}

A charged rotating black hole of mass $M$, charge $Q$ and angular momentum $J$ is described by the Kerr-Newman geometry (which in the Boyer-Lindquist coordinates $x^{\mu}=(t, r,\theta,\varphi)$ is given by):
\begin{equation}
ds^{2}=-c^2~(1-\frac{\Pi}{\Sigma})~ dt^{2}- 2~c ~\Pi ~L_{J}\sin^{2}\theta ~dt ~d\varphi +\frac{\Sigma}{\Lambda}~ dr^{2}+\Sigma 
~d\theta^{2}+ B  ~\frac{\sin^{2}\theta}{\Sigma}~ d\varphi^{2}
\label{eq:me}
\end{equation}
where 
\begin{equation}
\Pi=2 L_{c\ell}~ r - L^{2}_{Q}, ~~~\Sigma=r^{2}+ L_{J}^{2} \cos^{2}\theta, ~~~
~~\Lambda=r^{2} - \Pi + L_{J}^{2},~~~~B= (r^{2} + L_{J}^{2})^{2} - \Lambda  L_{J}^{2} \sin^{2}\theta
\label{eq:pi} 
\end{equation}
\begin{equation}
L_{c\ell}=\frac{GM}{c^{2}},~~~~L_{J}=\frac{J}{Mc},~~~~L_{Q}=\frac{\sqrt{G}Q}{c^{2}}
\label{eq:L}
\end{equation}
($G$ is the gravitational Newton constant), and $L_{c\ell}$, $L_{J}$, $L_{Q}$ satisfy at the classical level, the inequality :
\begin{equation}
L_{c\ell}^{2} \geq  L_{Q}^{2} +L_{J}^{2}
\label{eq:bcl}
\end{equation}

The equality holds for the extreme black hole, and Eq.~(\ref{eq:bcl}) shows the classical bound for $J$ and $Q$.  The two horizons are located at $r_{+}$ and $r_{-}$ :
\begin{equation}
r_{\pm}= L_{c\ell} \pm (L_{c\ell}^{2} -  L_{Q}^{2} - L_{J}^{2})^{1/2} 
\label{eq:rpm}
\end{equation}

The semiclassical black hole entropy is given by
\begin{equation}
S_{sem}^{(0)}= \frac{k_{B} A}{4 \ell^{2}_{Pl}} , ~~~~~~A= 4 \pi (r_{+}^{2} + L_{J}^{2}) 
\label{eq:Ssem}
\end{equation}
where $A$ is the black hole horizon surface, $k_{B}$ is the Boltzmann constant and $\ell_{Pl}$ the Planck length :
\begin{equation}
\ell_{Pl}= \Big(\frac{\hbar  G}{c^{3}} \Big)^{1/2} ~~~~~~~~.
\label{eq:lpl}
\end{equation}
This is the usual Bekenstein-Hawking entropy~\cite{6}\\
Inverting  Eq.(\ref{eq:Ssem}) yields the fundamental relation 
\begin{equation}
M = \Bigg \{  \Bigg[ \frac{1}{4\pi} \frac{S_{sem}^{(0)}}{\pi  k_{B}} m^{2}_{Pl} +\frac{\pi}{4} \frac{ k_{B}}{S} 
\bigg( \frac{M_{Q}^{2}}{m_{Pl}} \bigg)^{2} +\frac{M_{Q}^{2}}{2} \Bigg] \Bigg[  1- \pi \frac{k_{B}}{S}
\bigg(  \frac{M_{J}}{m_{Pl}} \bigg)^{2}  \Bigg]^{-1}\Bigg\}^{1/2}  
\label{eq:M}
\end{equation}
where 
\begin{equation}
M_{Q}=\frac{c^{2}L_Q}{G},~~~~M_{J}=\frac{c^{2}L_J}{G}, ~~~~m_{Pl}= \frac{c^{2}\ell_{Pl}}{G} 
\label{eq:MQJ}
\end{equation}
The semiclassical or QFT  black hole temperature (Hawking temperature)~\cite{7}, and the classical black hole angular velocity and electric potential are respectively: 
\begin{equation}
T_{sem}(J, Q) = \frac{\hbar c}{4 \pi k_{B}}~~ \frac{r_+ - r_-}{r_+^2 + L^2_J}, ~~~~
\Omega = \frac{c L_J}{r_+^2 + L^2_J}, ~~~~
\Phi = \frac{Q  r_+}{r_+^2 + L^2_J}
\label{eq:Tsem}
\end{equation}

$J=0$ and $Q=0$ yield the Schwarzschild black hole with 
\begin{equation}
r_+ \equiv  r_G = 2~ L_{cl},~~~~r_-= 0
\label{eq:rp}
\end{equation}
\section{QFT Semi classical entropy for the Kerr black hole\label{sec:SE}}

In the Quantum Field Theory (Q. F. T) description, ie semiclassical regime, the entropy and density of states of the Kerr black hole ($J\neq0$, $Q=0$) are related by
\begin{equation}
\rho_{sem}  = e^{\frac{S_{sem}(J)}{ k_B}} 
\label{eq:rsem}
\end{equation}

Eq.~(\ref{eq:Ssem}) expresses the zeroth order term $S_{sem}^{(0)}$ of the entropy in this regime. 
For $J\neq0$ and  $Q=0$,  Eq.~(\ref{eq:Ssem}) can be expressed as
\begin{equation}
S_{sem}^{(0)} (J)  =\frac{1}{2} (1+\Delta)  S_{sem}^{(0)}(J=0)    
\label{eq:S0}
\end{equation}
where
\begin{equation}
\Delta = \sqrt{1- \Bigg(  \frac{L_J}{L_{c\ell}} \Bigg)^2} = \sqrt{1- \Bigg( \frac{M_J}{M} \Bigg)^2} 
\label{eq:landa}
\end{equation}
and 
\begin{equation}
S_{sem}^{(0)} (J=0) = \frac{1}{2}  \frac{Mc^2}{T_{sem} (J=0)} 
\label{eq:SBH}
\end{equation}
\begin{equation}
T_{sem} (J=0) = \frac{\hbar c }{2 \pi k_B}  \frac{1}{4 L_{cl}} 
\label{eq:TBH}
\end{equation}

$S_{sem}^{(0)} (J=0)$ and $T_{sem} (J=0)$ are the entropy and the Hawking temperature of the Schwarzschild black hole respectively. $L_J$ and $M_J$ are defined by Eq.~(\ref{eq:L}) and Eq.~(\ref{eq:MQJ}) respectively.\\
Eq.~(\ref{eq:S0}) reads as well
\begin{equation}
S_{sem}^{(0)}(J)  =\frac{1}{2} \frac{ \mathcal{M} (J)c^2}{T_{sem}(J)} 
\label{eq:S0sem}
\end{equation}
where
\begin{equation}
\mathcal{M} (J) = M \Delta 
\label{eq:MJ}
\end{equation}
and $T_{sem}(J)$ is given by  Eq.~(\ref{eq:Tsem}) for $Q=0$ :
\begin{equation}
T_{sem}(J) = \frac{2 \Delta }{1 + \Delta} T_{sem}(J=0) 
\label{eq:TJ}
\end{equation}
$\Delta$ is given by Eq.~(\ref{eq:landa}). In terms of the length $ \mathit{L}_{c\ell} (J)$

\begin{equation}
\mathit{L}_{c\ell}(J) = 2 L_{c\ell} \Delta^{-1} (1 + \Delta),
\label{eq:LJ}
\end{equation}

Eq.~(\ref{eq:TJ}) is expressed as 
\begin{equation}
T_{sem}(J) = \frac{\hbar c }{2 \pi k_B}  \frac{1}{\mathit{L}_{c\ell}(J)}~~, 
\label{eq:TLJ}
\end{equation}
in analogy with Eq.~(\ref{eq:TBH}), ($\Delta = 1$ being the Schwarzschild case). 
\\

Eq.~(\ref{eq:S0}) and Eq.~(\ref{eq:TJ}) express the semiclassical entropy $ S_{sem}^{(0)}(J)$ and temperature $T_{sem}(J)$ of the Kerr black hole in terms of the corresponding quantities for $J=0$. $S_{sem}^{(0)}(J)$ can be thus expressed as a series expansion in $J$; or near an extremal black hole configuration $\Delta \ll 1$. 
The entropy $S_{sem}^{(0)}(J)$ is maximal for $ J=0$ and minimal for $\Delta = 0$ (the extremal black hole).
\\
The expression given by Eq.~(\ref{eq:SBH}), or Eq.~(\ref{eq:S0sem}), is just the {\it ordinary} expression of the entropy as for {\it any} ordinary physical system, where $T_{sem}$ is the Hawking temperature. This expression is appropriate to understand the analogy of black holes with other physical systems, astrophysical or particle systems, being them classical, quantum or strings. \\
The Hawking temperature $T_{sem}$ is {\it just} the Compton length of the black hole in the units of temperature. That is, the temperature scale of the semiclassical gravity properties. The {\it mass scale} of semiclassical gravity, $M_{sem}$, is precisely given by
\begin{equation}
T_{sem} = \frac{1}{8 \pi k_B} M_{sem} c^2~~, ~~~~~ M_{sem} = \frac {m_{Pl}^2}{M}
\label{eq:TMsem}
\end{equation}

In fact, the expression of $T_{sem}$ in terms of the surface gravity and the expression of $S_{sem}^{(0)}$ in terms of the area hold for {\it any} system in the semiclassical gravity regime. This semiclassical or intermediate energy regime interpolates between the classical and the quantum regimes of gravity. We discuss more on these regimes in Section~(\ref{sec:dual}). 
\\
On the other hand, and as we will see in Section~(\ref{sec:dual}), the usual Bekenstein-Hawking entropy $S_{sem}^{(0)}$ Eq.(\ref{eq:Ssem}), is just one term in a more general expression of the semiclassical entropy $S_{sem}$ which is a function of $S_{sem}^{(0)}$. For high mass, precisely  $M \gg m_{Pl}$, and  $J < GM^2/c$, $S_{sem}^{(0)}$ is the leading term. But for high angular momentum, (nearly extremal or extremal case $J = GM^2/c$), a gravitational phase transition operates and the whole Kerr entropy $S_{sem}$ is very different from $S_{sem}^{(0)}$. The whole Kerr black hole entropy $S_{sem}(J)$ Eq.~(\ref{eq:rsem}), as a function of the  Bekenstein-Hawking entropy $S_{sem}^{0}$, will be given in Section~(\ref{sec:dual}). 
\section{Quantum String Entropy in the Kerr background\label{sec:EQS}}

The string entropy $S_s(j)$ in a Kerr background is given, in terms of the string mass density of states $\rho_s(m, j)$, by
\begin{equation}
\rho_s(m,j)= e^{\frac{S_s(m,j)}{k_B}} 
\label{eq:rosj}
\end{equation}
where the string mode angular momentum $j$ is considered. The density of levels $d(n, j)$ of level $n$ and mode $j$ is the same in flat and curved space-time . The mass relation between $m$ and $n$ depends on the curvature of the background geometry and is in general different from the flat space-time string spectrum.
\\ 
The density of levels $d(n, j)$ ~\cite{8}, \cite{9} for large $n$ is given by, 
\begin{equation}
d(n, j) \sim n^{-a'} \Delta_s^{-2a'} e^{b \sqrt{n}~  \frac{1 + \Delta_s^2}
{2 \Delta_s }} 
\frac{1}{\cosh^2 \Bigg(  \frac{b}{4} \sqrt{n} ~\frac{(1 - \Delta_s^2)}{\Delta_s} \Bigg)} 
\label{eq:dnj}
\end{equation}
where 
\begin{equation}
\Delta_s(n, j) = \sqrt{1 -  \frac{j}{\hbar  n }}~~,~~~~~~ j\leq \hbar  n
\label{eq:landaj}
\end{equation}
The constants $a^{'}$ and $b$ depend only on the number of space-time dimensions and on the type of strings.\\
The string mass density of states $\rho_s(m, j)$ and the density of levels $d(n, j)$ are related by
\begin{equation}
\rho_s(\bar{m}, j/\hbar) ~d\bar{m} = d(n, j) dn 
\label{eq:rojd} 
\end{equation}
where
\begin{equation}
\bar{m} = \frac{m}{m_s}~~~~~,~~~~~ m_s = \sqrt{\frac{\hbar}{\alpha^{'} c}} ,
\label{eq:ms}
\end{equation}
$m_s$ being the fundamental string mass scale. The mass spectrum of strings in asymptotically flat spacetimes, as the black hole background, is the same as the flat spacetime string spectrum ~\cite{11}
\begin{equation}
\bar{m}^2 \simeq 4n ~~ \text{(closed)}  ~~~~,~~~~ \bar{m}^2 \simeq n ~~ \text{(open)}~~,~~n = 0,1,...
\label{eq:masyf}
\end{equation}
From Eqs.~(\ref{eq:dnj}) - (\ref{eq:masyf}) we have for closed strings
\begin{equation}
\rho_s(\bar{m}, j/\hbar) \sim  \bar{m}^{-a} \Delta_s^{-a-1} e^{ \frac{b}{2} \bar{m} 
\Big (\frac{1 +  \Delta_s ^2}{2 \Delta_s}\Big) }
\frac{1}{\cosh^2 \Bigg(\frac{b}{8} \bar{m} \frac{(1 - \Delta_s^2 )}{\Delta_s} \Bigg)} 
\label{eq:rojc}
\end{equation}
where
\begin{equation}
\Delta_s(m, j) = \sqrt{1 - \frac{4 j }{m^2 \alpha^{'} c}}~~,~~~4j\leq m^2 \alpha^{'} c~~~~, ~~~~a = D 
\label{eq:ba}
\end{equation}
and for open strings
\begin{equation}
\rho_s(\bar{m}, j/\hbar) \sim  \bar{m}^{-a} \Delta_s^{-a-1} e^{ b \bar{m} 
\Big (\frac {1 + \Delta_s^2}{2 \Delta_s}\Big)}
\frac{1}{\cosh^2 \Bigg(\frac{b}{4} \bar{m} \frac{(1 - \Delta_s^2)}{\Delta_s} \Bigg)} 
\label{eq:rojo}
\end{equation}
where 
\begin{equation}
\Delta_s(m, j) = \sqrt{1 - \frac{j }{m^2 \alpha^{'} c}}~~~~,~~~j\leq m^2 \alpha^{'} c~~,~~~~ a = (D-1)/2 
\label{eq:bao}
\end{equation}
being ~~$b = 2 \pi \sqrt{ \frac{D- 2 }{6}}$~~ and~~ $a  \equiv 2 a^{'} - 1$.
\\
From Eq.~(\ref{eq:rojc}), the asymptotic mass density of states can be written as
\begin{equation}
\rho_s(\bar{m}, j/\hbar) \sim  \rho_s(\bar{m}) \bar{F}(\bar{m}, j/\hbar) 
\label{eq:roF}
\end{equation}
where
\begin{equation}
\rho_s(\bar{m})  \sim  \bar{m}^{-a} e^{\frac{b}{2}\bar{m}}
\label{eq:ros}
\end{equation}
and 
\begin{equation}
\bar{F}(\bar{m}, j/\hbar) = \Delta_s^{-a-1} e^{ \frac{b}{4} \bar{m} 
\frac{1 +\Delta_s^2}{\Delta_s} }
\frac{1}{\cosh^2 \Bigg(\frac{b}{8} \bar{m} \frac{(1 - \Delta_s^2 )}{\Delta_s} \Bigg)}, 
\label{eq:F}
\end{equation}
$\bar{F}(\bar{m}, j/\hbar)$ takes into account the effect of the angular modes $j$, being $ \bar{F}(\bar{m},j=0) = 1$.\\ 
By introducing
\begin{equation}
S_s^{(0)} =   \frac{1}{2} \frac{mc^2}{T_s} ~~~\text{(closed~ strings)}, ~~~~
S_s^{(0)} =  \frac{mc^2}{T_s}~~~\text{(open~ strings)}, ~~~~
T_s =   \frac{m_s c^2}{k_B b} 
\label{eq:Ts}
\end{equation}
being $S_s^{(0)}$ the zero order string entropy for $j=0$ and $T_s$ the string temperature, both Eq.~(\ref{eq:rojc}) and Eq.~(\ref{eq:rojo}), can be expressed as
\begin{equation}
\rho_s(m, j)\sim \Big(\frac{S_s^{(0)}}{k_B}\Big)^{-a}  \Delta_s^{-a-1} e^{\Big(\frac {1 + \Delta_s^2}{2 \Delta_s}\Big)\frac{S_s^{(0)}}{k_B}}
\frac{1}{\cosh^2 \Bigg(\frac{S_s^{(0)}}{4k_B} \frac{(1 - \Delta_s^2)}{\Delta_s} \Bigg)} 
\label{eq:cos}
\end{equation}
or
\begin{equation}
\rho_s(m, j) \sim \Big(\frac{S_s^{(0)}}{k_B}\Big)^{-a} ~ e^{\Big(\frac{S_s^{(0)}}{k_B}\Big)} ~\bar{F}(S_s^{(0)}, j)~~,
\label{eq:roS0}
\end{equation}
with
\begin{equation}
\bar{F}(S_s^{(0)}, j) = \Delta_s^{-a-1} e^{\frac{S_s^{(0)}}{k_B} \frac{(1 - \Delta_s) ^2}{2 \Delta_s}} 
\frac{1}{\cosh^2 \Bigg(  \frac{S_s^{(0)}}{4 k_B} \frac{(1 - \Delta_s^2)}{\Delta_s} \Bigg)} 
\label{eq:Fcos}
\end{equation}

\begin{equation}
\Delta_s = \sqrt{1 - \frac{j}{\hbar} \Big(  \frac{k_B~b}{S_{s}^{(0)}} \Big)^2}
\label{eq:deltaSj}
\end{equation}

Therefore, from Eq.~(\ref{eq:rosj}), the string entropy $S_s(m,j)$ in the Kerr background is given by
\begin{equation}
S_s(m,j) = S_s^{(0)} - a ~k_B~\ln \Big (~\frac{S_s^{(0)}}{k_B}~ \Big)~+~k_B~\ln \bar{F}(S_s^{(0)}, j)      
 \label{eq:SjF}
\end{equation}
That is,
\begin{equation}
S_s(m,j) = \Big(\frac{1 + \Delta_s^2}{2 \Delta_s} \Big) S_s^{(0)} - a~k_B~\ln \Big (~\frac{S_s^{(0)}}{k_B}~ \Big) - 
(a + 1)~k_B~ \ln\Delta_s - 2~k_B~ \ln~\cosh \Big [~  \frac{S_s^{(0)}}{4 k_B}  \frac{(1 - \Delta_s^2)}{\Delta_s}~ \Big] 
\label{eq:SjFcos}
\end{equation}
being $S_s^{(0)}$ the zero order string entropy for $j=0$, Eq.~(\ref{eq:Ts}).\\ 
Notice the last term $k_B\ln \bar{F}(S_s^{(0)}, j)$ in Eq.~(\ref{eq:SjF}) which is enterely due to the angular momentum $j \neq 0$.\\ 
Interestingly enough, Eq.~(\ref{eq:SjFcos}) expresses the string entropy $S_s(m,j)$ for mass $m$ and spin $j$ in terms of the string entropy $S_s^{(0)}$ for $j=0$. The logarithmic terms have a negative sign. For $j=0$, we recover the expression:
\begin{equation}
S_s(m) = S_s^{(0)} - a ~k_B~\ln S_s^{(0)}
\label{eq:zero}
\end{equation}
In the first term of Eq.~(\ref{eq:SjFcos}), the factor in front of $S_s^{(0)}$ reads  is 
$ \Big[1 -\frac{j~ b}{2\hbar} \Big(\frac{k_B}{S_{s}^{(0)}}\Big)^2
\Big]/\Delta_s^{-1}$.\\
For $\Delta-s \neq 0$, the entropy $S_s(m,j)$ of string states of mass $m$ and mode $j$ is smaller than the string entropy for $j=0$. The effect of the spin is to reduce the entropy. $S_s(m,j)$ is maximal for $j=0$ (ie, for $\Delta_s = 1$).
\\
Is also instructive to express $S_s(m,j)$ in terms of the quantity $S_s^{(0)}(m, j)$ for $j\neq 0$ :
\begin{equation}
S_{s}^{(0)} (m,j)  =\frac{1}{2} (1+\Delta_s)  S_{s}^{(0)}    
\label{eq:Sosj}
\end{equation}
Then, $S_s(m,j)$ expresses as
\begin{equation}
S_s(m,j) = S_s^{(0)}(m,j) - a ~k_B~\ln \Big (~\frac{S_s^{(0)}(m,j)}{k_B}~ \Big)~+~k_B~\ln F(S_s^{(0)}, j) 
\label{eq:SjFs}
\end{equation}
with
\begin{equation}
F(S_s^{(0)},j)= \Big(\frac{1 + \Delta_s}{2}\Big)^{a}~e^{ \Big(\frac{1 -\Delta_s}{2}\Big) \frac{S_s^{(0)}}{k_B}}  ~~\bar{F}(S_s^{(0)},j) 
\label{eq:Fbar}
\end{equation}
\\
For $j=0$:~~~~~~$F = \bar{F} = 1$ ~~~~, ~~~~ $S_s^{(0)}(m,j=0)= S_s^{(0)}$.\\
Explicitely, in terms of the zero order entropy for $j\neq0$, $S_s^{(0)}(m,j)$:
\begin{equation}
F = \Delta_s^{-1} \Big(\frac{1 + \Delta_s}{2\Delta_s}\Big)^{a}~e^{\Big(\frac{1 -\Delta_s}{1 +\Delta_s}\Big) \frac{S_s^{(0)}(m, j)}{k_B \Delta_s}} 
\frac{1}{\cosh^2 \Bigg(  \frac{S_s^{(0)}(m, j)}{2 k_B} \frac{(1 - \Delta_s)}{\Delta_s} \Bigg)} 
\label{eq:Fstring}
\end{equation}
The argument of the last ($\ln\cosh $) term in Eq.~(\ref{eq:SjFcos}) is 
\begin{equation}
x ~\equiv~ \frac{S_s^{(0)}}{4 k_B}\frac{(1 - \Delta_s^2)}{\Delta_s}~=~\frac{b^2}{4\Delta_s}~
\frac {j}{\hbar}~\frac{k_B}{S_s^{(0)}}~ =~ \frac{b}{4\Delta_s}~ \frac{j}{\hbar}~\frac{m_s}{m}
\label{eq:argx}
\end{equation}
The last term in Eq.~(\ref{eq:SjFcos}) has the convergent series representation
\begin{equation}
\ln~\cosh x = - ~ \sum_{k=1}^{\infty}~ 2^{2k-1}~ \frac{(2^{2k}-1)} {k}~\frac{B_{2k}}{(2k)~!}~x^{2k} ~=~ -\frac{1}{2}~~ \sum_{k=1}^{\infty}~(-1)^k ~\frac{\sinh ^{2k}~x}{k}
\label{eq:lncosh}
\end{equation}
where $B_{2k} $ are the Bernouilli numbers.\\
For $j < (m/m_s)^2$, and $m \gg m_{s}$, that is for low $j$ and very excited string states, $S_s^{(0)}(m,j)$ is the leading term, but for high $j$, that is $j \rightarrow m^2 \alpha^{'} c$, ie $\Delta_s \rightarrow 0$, the situation is {\it very different} as we see it below.\\
Moreover, Eq.~(\ref{eq:SjFs}) for $S_s(m,j)$ allow us to write in Section~(\ref{sec:dual}) the whole expression for the semiclassical Kerr black hole entropy $S_{sem}$, as a function of the Bekenstein-Hawking entropy $S_{sem}^{(0)}$.
 
\section{Extremal String States: A New Transition \label{sec:EST}}

Let us notice the states in which $j$ reachs its {\bf maximal} value, that is $j =  m^2 \alpha^{'} c$ , we call these states {\bf "extremal string states"}. In this case, $\Delta_s = 0$, and the term $S_s^{(0)}(m,j)$ is minimal:
\begin{equation}
S_{s}^{(0)} (m,j)_{extremal}  = \frac{1}{2} ~S_{s}^{(0)}    
\label{eq:Sosextrem}
\end{equation}
But for $\Delta_s \rightarrow 0$, the last term in Eq.~(\ref{eq:SjFs}) substracts the first one, the pole in $\Delta_s$ of these two terms cancel out, and thus does $S_s^{(0)}$, yielding :
\begin{equation} 
S_s(m,j)_{extremal} =   -(a + 1)~k_B~ \ln \frac {\Delta_s}{2}~ +~ k_B~\ln 2~ +~ \Delta_s ~\Big(~\frac{3} {4}~S_s^{(0)}~-~ak_B \Big)~+ ~ O (\Delta_s^{2})
\label{eq:Ssextremal}
\end{equation}
In terms of the mass, or temperature:
\begin{equation} 
\Delta_s= \sqrt{1 - \frac{j}{\hbar} \Big(\frac{m_s}{m} \Big)^2} =  \sqrt{1 - \frac{j}{\hbar} \Big(\frac{T_s}{T} \Big)^2} ~~,~~ T = \frac{m c^2}{k_B b}, 
\label{eq:deltamass}
\end{equation}
Thus, in the extreme limit $(j/\hbar) \rightarrow (m/m_s)^2 $ :
\begin{equation} 
\Delta_{s~extremal} = \sqrt{\frac{2}{m}}~\sqrt{m - \Big(\frac{j}{\hbar}\Big)^{1/2} {m_s}}~=~\sqrt{\frac{2}{T}}~\sqrt{T - \Big(\frac{j}{\hbar}\Big)^{1/2} {T_s}}
\label{eq:deltasextremal}
\end{equation}
and $S_s(m,j)_{extremal}$ is dominated by
\begin{equation}
S_s(m,j)_{extremal} =  -(a + 1)~k_B~ \ln~ (~ \sqrt{\frac{2}{T}}\sqrt{T - \Big(\frac{j}{\hbar}\Big)^{1/2} {T_s}}~)~+~O(1)
\label{eq:STextremal}
\end{equation}
This shows that a {\bf phase transition} takes place at $T \rightarrow \sqrt{(j/\hbar)}~T_s$, we call it {\bf extremal} transition. Notice that this is {\bf not} the usual (Hagedorn/Carlitz) string phase transition occuring for $m \rightarrow \infty $, $T \rightarrow T_s$; such transition is also present for $j\neq 0$ since $\rho (m,j)$ has the same $m \rightarrow \infty $ behaviour as $\rho(m)$.\\
The extremal transition we found here is a {\bf gravitational} like phase transition: the square root {\it branch point} behaviour near the transition is analogous to that found in the thermal self-gravitating gas of (non-relativistic) particles (by mean field and Monte Carlo methods)\cite {5}. And this is also the same behaviour found for the microscopic density of states and entropy of strings in de Sitter background \cite{4}. \\
As pointed out in \cite {4}, this string behaviour is {\it universal}: the logarithmic singularity in the entropy characteristic of this transition (or pole singularity in the specific heat) holds in any number of dimensions, and is due to the gravitational interaction in the presence of temperature, similar to the Jeans's like instability at finite temperature but with a more complex structure.
\\
A particular new aspect here is that the transition shows up at high angular momentum, (while in the thermal gravitational gaz or for strings in de Sitter space, angular momentum was not considered, (although it could be taken into account)). 
\\
Since $j\neq 0$, the extremal transition occurs at a temperature  $ T_{sj}~=~ \sqrt{j/\hbar}~T_s $, {\it higher} than the string temperature $T_{s}$.
That is, angular momentum which acts in the sense of the string tension, appears in the transition as an "effective string tension" : a smaller  $\alpha^{'}_j~  \equiv ~ \sqrt{\hbar/j}~ \alpha^{'}$ ~ (and thus a {\it higher tension}).       
 
These new results desserve more investigation and we will not extend more on them here.

\section{Reissner-Nordstr\"om black hole: QFT and Strings\label{sec:RNBH}}

In the Q.F.T. (or semi classical) regime, the entropy and density of states of the Reissner-Nordstr\"om black hole $(Q \neq~ 0, J=0 )$ are related by
\begin{equation}
\rho_{sem} = e^{\frac{S_{sem}(Q)}{k_B}}
\label{eq:roq}
\end{equation}
As before, Eq.~(\ref{eq:Ssem}) expresses the zeroth order term of the entropy in this regime, (ie, this is the usual Bekenstein- Hawking entropy). For $J=0$ and $Q \neq 0$,  Eq.~(\ref{eq:Ssem}) becomes
\begin{equation}
S_{sem}^{(0)}(Q) = \Big[ \frac{1}{2} (1 + \widehat{\Delta}) \Big]^2~~ S_{sem}^{(0)}(Q=0) 
\label{eq:S0q}
\end{equation}
where $S_{sem}^{(0)}(Q=0)$ (Schwarzschild black hole entropy), is given by Eq.~(\ref{eq:SBH}) and
\begin{equation}
 \widehat{\Delta} = \sqrt{1-\Big( \frac{L_Q}{L_{c\ell}}\Big)^2} = \sqrt{1-\Big( \frac{M_Q}{M}\Big)^2}
\label{eq:lanhat}
\end{equation}
$L_Q$ and $M_Q$ being defined by Eqs.~(\ref{eq:L}) and ~(\ref{eq:MQJ}).  $S_{sem}^{(0)}(Q)$ can be written

\begin{equation}
S_{sem}^{(0)}(Q) = \frac{1}{2}  \frac{\mathcal{M} (Q)~c^{2}} {T_{sem}(Q)} 
\label{eq:S0qM}
\end{equation}
where
\begin{equation}
\mathcal{M}(Q) = M \widehat{\Delta} 
\label{eq:Mq}
\end{equation}
and
\begin{equation}
T_{sem}(Q) = \frac{\widehat{\Delta}}{\Big[ \frac{1}{2} (1 +\widehat{\Delta})\Big]^2}  ~~T_{sem}(Q=0) 
\label{eq:Tq}
\end{equation}
being $T_{sem}(Q=0)$ the Hawking temperature of the Schwarzschild black hole Eq.~(\ref{eq:TBH}).
In terms of the length $L_{c\ell}(Q)$
\begin{equation}
L_{c\ell}(Q) = L_{c\ell} ~\widehat{\Delta}^{-1} ~(1 + \widehat{\Delta})^2
\label{eq:Lq}
\end{equation}
Eq.~(\ref{eq:Tq}) becomes
\begin{equation}
T_{sem}(Q) = \frac{\hbar c }{2 \pi k_B} ~ \frac{1}{L_{c\ell}(Q)} 
\label{eq:TqL}
\end{equation}
in analogy with Eqs.~(\ref{eq:TBH}) and  (\ref{eq:TLJ}). \\
On the other hand, the entropy of strings in a Reissner-Nordstr\"om background is given by
\begin{equation}
\rho_s(m, q)= e^{\frac{S_s(m,q)}{k_B}}  
\label{eq:roqS}
\end{equation}
where $\rho_s(m, q)$ is the string density of states of mass $m$ and charge mode  $q$.\\
For large $m$, $\rho_s(m, q)$ ~\cite{10} ($D=4$) is given by

\begin{equation}
\rho_s(m, q) \sim  \rho_s(m) ~\exp \Big\{\frac{-q^2}{\hbar c (m/m_s)} \Big\}  
\label{eq:romsq}
\end{equation}

Eqs.~((\ref{eq:roqS}) - (\ref{eq:romsq})) yield: 
\begin{equation}
S_s(m,q)  =  S_s(m) - \frac{k_B}{\hbar c}  \frac{b~q^2}{S_{s}^{(0)}} =  S_s^{(0)} - a ~k_B~\ln S_s^{(0)}
 - \frac{k_B}{\hbar c}  \frac{b~q^2}{S_{s}^{(0)}}
\label{eq:Sqb}
\end{equation}
where $S_s(m)$ is the entropy for $q=0$ Eq.~(\ref{eq:zero}), and $S_{s}^{(0)}$, its leading term, is given by Eq.~(\ref{eq:Ts}).
\\  
The string entropy $S_s(m,q)$ of mass $m$ and mode charge $q$ is smaller than the entropy for $q = 0$. As the effect of the spin mode $j$, the effect of the charge $q$ is to reduce the entropy; the $q$-reduction is proportional to $q^{2}$, while the $j$-reduction to the entropy is linear in $j$ plus logarithmic corrections.


\section{Partition Function of strings in the Kerr-Newman background\label{sec:partition}}

The Kerr-Newman background is asymptotically flat. Black hole evaporation -and "any slow-down" of this process- will be measured by an observer which is at this asymptotic region. In this region, the thermodynamical behaviour of the string states is deduced from the canonical partition function: 
\begin{equation}
\ln Z = \frac{V_{D-1}}{(2 \pi)^{D-1}} \sum_{j, \alpha} \int^{\infty}_{m_0} d \Big( \frac{m}{m_s}\Big) \rho_s(m, j, q) 
\int d^{(D-1)}k 
 \ln \Bigg\{\frac{1 + \exp \Big\{-\beta_{sem} [(m^2 c^4 + \hbar^2 k^2 c^2)^{\frac{1}{2}}-\mu_{\alpha}]\Big\}}
{1 - \exp \Big\{-\beta_{sem} [(m^2 c^4 + \hbar^2 k^2 c^2)^{\frac{1}{2}}-\mu_{\alpha}]\Big\}}\Bigg\} 
\label{eq:Z}
\end{equation}
where $\beta_{sem}=(k_B T_{sem})^{-1}$, $T_{sem}$ is given by Eq.~(\ref{eq:Tsem}); $\mu_{\alpha}$,  $j_{\alpha}$ and  $q_{\alpha}$ are the chemical potential, angular momentum and charge of a string mode $\alpha$ respectively; $m_0$  is the lowest string mass for which the asymptotic string density of mass level is valid, and $m_{s}$ is the fundamental string mass scale Eq.~(\ref{eq:ms}).  

The chemical potential $\mu_{\alpha}$ is given by
\begin{equation}
\mu_{\alpha} = j_{\alpha}\Omega + q_{\alpha}\Phi 
\label{eq:mu}
\end{equation}
where $\Omega$ and $\Phi$ are the Kerr-Newman angular velocity and electric potential respectively. The string mass
density of states  $\rho_s(m, j, q)$ is given by Eqs.~(\ref{eq:rojc}),~(\ref{eq:rojo}), and~(\ref{eq:romsq});  $j_{\alpha}$ is the string mode angular momentum about the axis of rotation of the black hole  ($j_{\alpha}= n_{\alpha} \hbar$). 

From Eq.~(\ref{eq:Z}) we have
\begin{equation}
\ln Z = \frac{4 V_{D-1}}{(2 \pi)^{D/2}} \frac{c}{\beta_{sem}^{\frac{D-2}{2}}\hbar^{(D-1)}}  
\sum_{j, \alpha} \sum_{n=1}^{\infty} \frac{e^{(2n-1) \beta_{sem} \mu_{\alpha}}}{(2n-1)^{D/2}}  
\int^{\infty}_{m_0} d \Big(\frac{m}{m_s}\Big) \rho_s (m, j, q)  m^{D/2} K_{D/2} \big[ \beta_{sem} (2n-1) m c^2 \big]
\label{eq:lnZ}
\end{equation}
$K_{D/2}$ being the modified Bessel function. Considering the asymptotic behaviour of the string mass density of states \begin{equation}
\rho_s \sim \Big(\frac{m}{m_s}\Big)^{-a}~e^{-\beta_{s}mc^{2}},
\label{eq:rhoasy}
\end{equation}
and the leading order ($\beta_{sem}~ mc^2 \gg1$):
\begin{equation}
K_{\nu} \sim_{(z \rightarrow \infty)} \Big( \frac{\pi}{2 z} \Big)^{1/2} e^{-z}
\label{eq:Knu},
\end{equation}
 we obtain
\begin{equation}
\ln Z \simeq \frac{2 V_{D-1}}{(2 \pi)^{\frac{D-1}{2}}} \frac{1}{ (\beta_{sem} \hbar^2)^{\frac{D-1}{2}}}  
\sum_{j,  \alpha} e^{\beta_{sem} \mu_{\alpha}} 
\int^{\infty}_{m_0} d \Big(\frac{m}{m_s}\Big)~ \rho_s(m, j, q)~  m^{\frac{D-1}{2}} e^{-\beta_{sem} mc^2}
\label{eq:Zjq}
\end{equation}
Then, the leading order contribution to $\ln Z$ for large $m$ ($m > m_0$) is :
\\
\begin{equation}
\ln Z \simeq \frac{2 V_{D-1}} {(2 \pi)^{\frac{D-1}{2}}} \frac{m_s^{a-1}} { (\beta_{sem} \hbar^2)^{\frac{D-1}{2}}}  
\sum_{\alpha} e^{\beta_{sem} \mu_{\alpha}}
\int^{\infty}_{m_0} dm~ m^{(-a + \frac{D-1}{2})}~ e^{-( \beta_{sem}-\beta_{s}) mc^2}
\label{eq:Zaprox}
\end{equation}
\\
where $\beta_{s} = (k_B T_{s})^{-1}$ Eq.~(\ref{eq:Ts}). The factor $2$ in front of Eq.~(\ref{eq:Zaprox}) stands for both bosonic and fermionic strings included; (otherwise, this factor is absent for either bosonic or fermionic strings).
Eq.~(\ref{eq:Zaprox}) implies that the black hole temperature $T_{sem}$ is bounded by the string temperature $T_{s}$: $T_{sem}(J, Q) \leq T_{s}$. We discuss this bound in Section III below. \\
Finally,
\begin{equation}
\ln Z = \frac{2 V_{D-1}} {(2 \pi)^{\frac{D-1}{2}}} \frac{m_s^{a-1}} { (\beta_{sem} \hbar^2)^{\frac{D-1}{2}}}  
\sum_{\alpha} e^{\beta_{sem} \mu_{\alpha}}
\frac{1} {\Big[ (\beta_{sem} - \beta_{s})c^2 \Big]^{ -a + \frac{D+1}{2}}} 
~~\Gamma~\Big(-a + \frac{D+1}{2}, m_0(\beta_{sem} - \beta_{s})c^2  \Big)
\label{eq:Zgama}
\end{equation}
where $\Gamma(\alpha, x)$ is the incomplete gamma function. For low temperature ($T_{sem}\ll T_{s}$) and high temperature ($T_{sem}\rightarrow T_{s}$), ln Z (open strings, $a=\frac{D-1}{2}$), behaves as: 

 For $\beta_{sem} \gg\beta_{s}$:
\begin{equation}
\ln Z \simeq V_{D-1} \Big( \frac{m_s}{2 \pi \beta_{sem} \hbar^2} \Big)^{\frac{D-1}{2}} 
\sum_{\alpha} e^{\beta_{sem} \mu_{\alpha}}~~ e^{-\beta_{sem}m_0 c^2}
\label{eq:TTs}
\end{equation}

 For $\beta_{sem} \rightarrow \beta_{s}$:
\begin{equation}
\ln Z \simeq V_{D-1} \Big( \frac{m_s}{2 \pi \beta_{s} \hbar^2} \Big)^{\frac{D-1}{2}} 
\sum_{\alpha} e^{\beta_{sem} \mu_{\alpha}} \frac{1} {\Big[(\beta_{sem} - \beta_{s})m_s c^2 \Big]}
\label{eq:Zo}
\end{equation}
This shows an universal pole singularity for any $D$ at the temperature $T_s$, typical of a string system with intrinsic Hagedorn temperature, and indicates a string phase transition of Carlitz's type~\cite{12} to a condensate finite energy state. Here, the transition takes place at $T_{sem}$ = $T_{s}$ towards a microscopic finite energy condensate of size $L_{s}$. This stringy state forms at the last stage of black hole evaporation, from the massive very excited strings emitted by the black hole, as can be seen precisely from the string emission cross section computed in Section (\ref{sec:qe}) below.
\section{Quantum String Emission by Kerr Newman black holes\label{sec:qe}}

In the string analog model, or thermodynamical approach, the quantum string emission by the black hole is given by the cross section
\begin{equation}
\sigma_{string} = \sum_{j} \sum_{\alpha} \int_{m_0}^{\infty} \rho_{s}(m, j, \alpha)~\sigma_{\alpha}(m,j)~  
d\big( \frac{m}{m_s}\big) 
\label{eq:sD}
\end{equation}
where 
\begin{equation}
\sigma_{\alpha}(m, j) = \int_{0}^{\infty} \sigma_{\alpha}(k,  j)~ d\mu(k, j)
\label{eq:sDj}
\end{equation}
$\sigma_{\alpha}(k,  j)$ is the QFT  
emission cross section of particles of species ${\alpha}$ in a mode of frequency $k$, spin $j_{\alpha}$ = $n_{\alpha}\hbar$, and charge $q_{\alpha}$:

\begin{equation}
\sigma_{\alpha}(k, j) = \frac{\Gamma_{\alpha}}{\exp \Big\{ \beta_{sem}[E(k) - \mu_{\alpha}] \Big\} + (-1)^{2j +1}}
\label{eq:sk}
\end{equation}
$\Gamma_{\alpha}$ is the classical absorption cross section (grey body factor), and $\mu_{\alpha}$ the chemical potential given by Eq.~(\ref{eq:mu}).

The QFT  emission cross section of particles of mass $m$ is defined as
\begin{equation}
\sigma(m) = \sum_{j} \sum_{\alpha} \int_{0}^{\infty} \sigma_{\alpha}(k, j)~  d\mu(k, j) 
\label{eq:sm}
\end{equation}
where $d\mu(k,j)$ is the number of states between $k$ and $k+dk$
\begin{equation}
d\mu(k, j) = n_{j} \frac{2 V_{D-1}}{(4 \pi)^{\frac{D-1}{2}}}~\frac{k^{D-2}}{\Gamma \Big( \frac{D-1}{2} \Big)} ~dk
\label{eq:muj}
\end{equation}
and $n_{j}$ is the number of spin states : $[( j+ D -3)! (2j + D -2)]~ [j! (D- 2)!]^{-1}$ for $SO(D)$; ( $2j+1$ for $SO(3)$).

From Eq.~(\ref{eq:sm}) we have
\begin{eqnarray}
&&\sigma(m)=\sum_{j} \sum_{\alpha}n_{j} \frac{V_{D-1}}{(2 \pi)^{\frac{D-1}{2}}}
\frac{\Gamma_{\alpha}~~\Big( mc^2\Big)^{\frac{D-2}{2}}}{\beta_{sem}^{D/2}~ (\hbar c)^{D-1}} ~~
\times \\
&&\sqrt{\frac{2}{\pi}}~\sum_{n=1}^{\infty} \frac{(-1)^{(n-1)2j}}{n^{D/2}}~e^{n \mu_{\alpha}\beta}
\Big\{ n\beta_{sem} mc^2 K_{_{D/2}} (n\beta_{sem} mc^2) + K_{_{D/2 - 1}}  (n\beta_{sem} mc^2) \Big\} 
\label{eq:smD}
\end{eqnarray}
$K_{_{D/2}}$ being the modified Bessel function. For large $m$ and leading order $(n=1)$ with the asymptotic behaviour Eq.~(\ref{eq:Knu}) we have :
\begin{equation}
\sigma(m)\simeq \sum_{j} \sum_{\alpha}n_{j}\frac{V_{D-1}}{(2 \pi)^{\frac{D-1}{2}}}
\frac{ \Gamma_{\alpha}~~\Big( mc^2\Big)^{\frac{D-3}{2}} }{\beta_{sem}^{\frac{D+1}{2}}~ (\hbar c)^{D-1}}
e^{-\beta_{sem}(mc^2-\mu_{\alpha})}
\label{eq:smDj}
\end{equation}
Therefore, for the quantum emission of strings by the black hole, $\sigma_{string}$ Eq.~(\ref{eq:sD}), we find
\begin{equation}
\sigma_{string}\simeq\sum_{j} \sum_{\alpha}n_{j}\frac{V_{D-1}}{(2 \pi)^{\frac{D-1}{2}}}
\frac {\Gamma_{\alpha}~~c^{D-3}}{\beta_{sem}^{\frac{D+1}{2}}(\hbar c)^{D-1}}
\frac{e^{\beta_{sem}\mu_{\alpha}}}{(m_s)^{-a+1}}\int^{\infty}_{m_0}m^{-a+\frac{D-3}{2}}
e^{-(\beta_{sem}-\beta_{s})mc^2}dm
\label{eq:sDap}
\end{equation}

which leads to

\begin{equation}
\sigma_{string} \simeq\sum_{j} \sum_{\alpha}n_{j}\frac{V_{D-1}}{(2 \pi)^{\frac{D-1}{2}}}~
\frac{e^{\beta_{sem} \mu_{\alpha}}~~\Gamma_{\alpha}~~m_s^{\frac{D-3}{2}}} {(\beta_{sem}~\hbar^{2}c^2)^{\frac{D-1}{2}}}~
\frac{ e^{-(\beta_{sem}-\beta_{s})m_0 c^2}}{\left(\beta_{sem}-\beta_{s}\right)}
 \label{eq:smia}
\end{equation}

For $\beta_{sem}\gg \beta_{s}$ :
\begin{equation}
\sigma_{string}\simeq \sum_{j} \sum_{\alpha}n_{j}\frac{V_{D-1}}{(2 \pi)^{\frac{D-1}{2}}}~
\frac{\Gamma_{\alpha}~~m_s^{\frac{D-3}{2}}}{\beta_{sem}^{\frac{D+1}{2}}~ (\hbar c)^{D-1}}~
e^{\beta_{sem} \mu_{\alpha}}~e^{-\beta_{sem}m_0c^2}
\end{equation}

For $\beta_{sem} \rightarrow \beta_{s}$ :
\begin{equation}
\sigma_{string}\simeq \sum_{j} \sum_{\alpha}n_{j}\frac{V_{D-1}}{(2 \pi)^{\frac{D-1}{2}}}~
\frac{\Gamma_{\alpha}~~m_s^{\frac{D-3}{2}}}{\beta_{s}^{\frac{D-1}{2}}~ (\hbar c)^{D-1}}~
e^{\beta_{s} \mu_{\alpha}} \frac{1}{(\beta_{sem} - \beta_s)}
\label{eq:sDTs}
\end{equation}
\\
The string emission cross section shows that for $T_{sem}\ll T_{s}$, the emission is thermal with temperature $T_{sem}$, this is the Hawking part of the emission, in the early stage of evaporation, that is, the semiclassical or QFT regime. As evaporation proceeds, $T_{sem}$ increases:
for $T_{sem}\rightarrow T_{s}$, the massive string modes dominate the emission, the string emission cross section shows at $T_{s}$ the same behaviour as ln$Z$ Eq.~(\ref{eq:Zo}), that is, a phase transition takes place at $T_{sem}$ = $T_{s}$. This phase transition undergone by the emitted strings represents the non-perturbative back reaction effect of the string emission on the black hole. An explicit dynamical perturbative solution to the back reaction effect of the string emission on the  Schwarzschild black hole ~\cite{2} accompasses this picture: the black hole losses its mass, reduces its radius until $L_{s}$ and rises its temperature until $T_{s}$, in a non-singular finite process, the temperature does not becomes infinite but remains bounded by $T_{s}$ (and the radius and mass, do not reduce to zero).\\
Moreover, in the semiclassical (QFT) regime, ie in the early stages of black hole evaporation, the black hole decays as a grey body at the Hawking temperature $T_{sem}$, with decay rate
\begin{equation}
\Gamma_{sem} = \left| \frac{ d\ln M_{sem} }{ d t}\right| \sim G~T_{sem}^{3}~~,~~M_{sem}= 8~\pi~ (\hbar=c=k_{B}=1).           
\label{eq:decay}
\end{equation}
As evaporation proceeds, $T_{sem}$ increases until it reaches the string temperature $T_{s}$, the black hole itself becomes a very excited string state. This quantum string state decays in the usual way quantum strings do~ \cite{4},\cite{13}, ie with a width, 
\begin{equation}
\Gamma_{s}\sim \alpha'~T_{s}^{3}           
\label{eq:des}
\end{equation}
As $T_{sem} \rightarrow  T_s$, $\Gamma_{sem}$ becomes $\Gamma_{s}$, ($G\sim \alpha'$),  the final decay is a pure (non mixed) quantum mechanical string decay into all type of particles.

\section{String bounds on the black hole temperature, angular momentum and charge\label{sec:bTJQ}}

The Kerr Newman black hole temperature in the semiclassical (QFT) regime Eq.~(\ref{eq:Tsem}) can be expressed
\begin{equation}
T_{sem}(J, Q) = \frac{\hbar c }{2 \pi k_B}  \frac{1}{L_{c\ell}(J, Q)} 
\label{eq:TsemJQ}
\end{equation}
where
\begin{equation}
L_{c\ell} (J, Q) = \frac{2~L_{c\ell}}{\delta} \left(1~+~\delta~-~\frac{\nu^{2}}{2}\right),
\label{eq:LJQ}
\end{equation}
\begin{equation}
\nu\equiv \frac{L_{Q}}{L_{c\ell}}~~~~~~,~~~~~~\mu\equiv\frac{L_{J}}{L_{c\ell}}~~~~~~,~~~~~~ \delta\equiv \sqrt{1~-~\nu^{2}~-~\mu^{2}}
\label{eq:numu}
\end{equation}
with $L_{c\ell}$, $L_{J}$ and  $L_{Q}$ given by Eq.~(\ref{eq:L}).
From Eq.~(\ref{eq:bcl}), we have
\begin{equation}
\nu^{2}~+~\mu^{2}\leq 1
\label{eq:nucu}
\end{equation}
and furthermore
\begin{equation}
\mu^{2}\leq 1~~~~~\text{and}~~~~~\nu^{2}\leq 1
\label{eq:mucu}
\end{equation}
That is, in the semiclassical (Q.F.T) regime, one always has
\begin{equation}
T_{sem}(J, Q) \leq  T_{sem}(J=0=Q)~~,
\label{eq:Tclb}
\end{equation}
which implies
\begin{equation}
L_{c\ell}(J, Q)\geq 4~L_{c\ell}
\label{eq:Lclb}
\end{equation}
The equality
\begin{equation}
\nu^{2}~+~\mu^{2} = 1
\label{eq:nuex}
\end{equation}
corresponding to the extremal Kerr Newman black hole, $T_{sem}(J, Q)_{extremal} = 0$. \\
On the other hand, in the string regime, from the canonical partition function Eq.~(\ref{eq:Zaprox}), we see that the Kerr Newman black hole temperature Eq.~(\ref{eq:Tsem}) is bounded by the string temperature $T_{s}$ Eq.~(\ref{eq:Ts}). $T_{s}$ can be written as 
\begin{equation}
T_{s} = \frac{\hbar c }{2 \pi k_B}  \frac{1}{L_{s}}
\label{eq:TLs}
\end{equation}
with the  string length
\begin{equation}
L_{s} = \frac{b}{2 \pi}\sqrt{\frac{\hbar \alpha'}{c}}~. 
\label{eq:Ls}
\end{equation}
Therefore, the  string bound 
\begin{equation}
T_{sem}(J, Q) \leq T_{s}
\label{eq:Tsb}
\end{equation}
implies
\begin{equation}
L_{c\ell}(J, Q) \geq L_{s}.
\label{eq:Lsb}
\end{equation}
Eq.~(\ref{eq:LJQ}) and Eq.~(\ref{eq:Lsb}) yield
\begin{equation}
1~-~\frac{\nu^{2}}{2}\geq \delta~\sigma,~~~~~~~~
\sigma \equiv \frac{L_{s}}{2~L_{c\ell}}~-~1
\label{eq:nub}
\end{equation}
The above relations, Eq.~(\ref{eq:Lsb}) and  Eq.~(\ref{eq:nub}) lead to three different situations:

(i) $ \sigma < 1$ (i.e. $L_{c\ell} > L_{s}/4$). In this case, the inequality  Eq.~(\ref{eq:nub}) is satisfied for all nonvanishing values of $\mu$ and $\nu$.

(ii) $ \sigma = 1$ (i.e. $L_{c\ell} = L_{s}/4$). One has $\mu=0$ = $\nu$ (with $L_{c\ell}(J=0=Q)~=~ L_{s}$) for the equal sign in Eq.~(\ref{eq:nub}); and $\mu\neq 0$, $\nu\neq 0$ or ($\mu\neq 0$, $\nu$= $0$), or ($\mu = 0$, $\nu\neq 0$) for the strict inequality. 

(iii) $ \sigma > 1$ (i.e. $L_{c\ell} < L_{s}/4$). In this case, two critical values, $ \mu_{0}$  and $\nu_{0}$, for the angular momentum parameter $\mu$ and the charge parameter $\nu$ respectively, appear. They are given by
\begin{equation}
\mu_{0}^{2} = \frac{4~\left(\sigma^{2}-1\right)\left(1-\nu^{2}\right)~-~\nu^{4}}{4~\sigma^{2}}
\label{eq:muo}
\end{equation}
and
\begin{equation}
\nu_{0}^{2} = 2~\left( 1~-~\sigma^{2}~+~\sigma\sqrt{\sigma^{2}~-~1}\right)
\label{eq:nuo}
\end{equation}

If $\nu\geq\nu_{0}$, Eq.~(\ref{eq:nub}) is fulfilled for all $\mu's$ and saturated for $\nu=\nu_{0}$ and $\mu=0$. In the opposite case i.e. $\nu<\nu_{0}$, one must have $\mu\geq\mu_{0}$. Here, $L_{c\ell}(J,Q) = L_{s}$ if $\mu = \mu_{0}$.
\\
Let us summarize the above discussion, and translate it in terms of masses and temperatures. With the help of Eq.~(\ref{eq:L}) and the mass scale, $M_{s} = \frac{c^{2}}{G} L_{s}$,
the three cases above express as:

(i) If $M > M_{s}/4$, then $T_{sem}(J,Q) < T_{s}$ is always verified without restrictions on $J$ and $Q$.

(ii) If $M~ =~ M_{s}/4$, then $T_{sem}(J,Q) < T_{s}$ with any value of $J$ and $Q$ excluding both J and Q simultaneously equal to zero. The string limit $T_{sem}(J,Q) = T_{s}$ is reached with both $J = 0$ and $Q = 0$. 

(iii) If $M < M_{s}/4$, there  exist a critical value for the charge, $Q_{0}$, and a critical value for the angular momentum, $J_{o}$:  For $Q\geq Q_{0}$, $T_{sem}(J, Q)\leq T_{s}$ holds for all $J$; the string limit $T_{sem}(J, Q)$ = $T_{s}$ is reached for $Q =Q_{0}$ and $J=0$ ; $Q_{0}$ is given by
\begin{equation}
	Q_{0}^{2} =  \frac{G}{2}
	                            \left( 4 M^{2}~-~\left( M_{s} - 2M \right)^{2}
	                                    + \left( M_{s} - 2M\right) \sqrt{ \left(M_{s}~-~2M\right)^{2} - 4M^{2}} \right)
\label{eq:Q0}
\end{equation}

Otherwise, namely, if $Q < Q_{o}$, there is a minimal angular momentum, $ J\geq J_{o}$, given by:
\begin{equation}
	J_{0}^{2} =  \frac{ \frac{ G^{2} M^{4} }{ c^{2} }
			             \left( \left(M_{s}-2M\right)^{2} - 4M^{2} \right)
			             \left(1-\frac{ Q^{2} }{ GM^{2} }\right)
			             - \frac{ Q^{4}~M^{2} }{ c^{2} }                                                                   
			         }{\left(M_{s}~-~2M\right)^{2}}
\label{eq:J0}
\end{equation}
Here, the limit $T_{sem}(J,Q) = T_{s}$ is reached for $J =J_{0}$.
\\
From this analysis we can conclude that  -  given a semiclassical Kerr Newman black hole with mass $M$, angular momentum $J$, charge $Q$ and temperature $T_{sem}(J,Q)~<~ T_{s}$  - there are three possible cases for its evolution into a string state with temperature $T_{s}$:

(I) If $M > M_{s}/4$, $J\neq 0$ and $Q\neq0$, the semiclassical Kerr Newman black hole reaches a string state of temperature $T_{s}$, mass $M = M_{s}/4$, angular momentum  $J = 0$ and charge $Q = 0$.

(II) If $M < M_{s}/4$, $J\neq0$ and $Q > Q_{0}$ ($Q_{0}$ being the critical value Eq.~(\ref {eq:Q0})), the string temperature $T_{s}$ is reached with $J=0$ and $Q=Q_{0}$.

(III) If $M < M_{s}/4$, $J >J_{0}$ and $Q < Q_{0}$, the string temperature $T_{s}$ is reached for $J=J_{0}$, being $J_{0}$ the minimal value Eq.~(\ref {eq:J0}). In this string state both angular momentum and charge different from zero. 
\\
Besides the classical bounds implying a maximal angular momentum and maximal charge for the black hole, there is a {\it minimal} value for the angular momentum and the charge in the string regime. 
\section{The extremal black hole\label{sec:EBH}}

The semiclassical extremal Kerr Newman black hole does not evaporate through Hawking radiation, the Hawking temperature is zero in this case (Eqs.~(\ref{eq:TsemJQ}),~(\ref{eq:LJQ}) and~(\ref{eq:nuex})) 
\begin{equation}
T_{sem}(J, Q)_{extremal} = 0
\label{eq:Textremal } 
\end{equation}
Eq.~(\ref{eq:Tsb}) is a strict inequality in this case. The string temperature can not be reached (unless the extremal configuration would be already a stringy state).
\\
The extremal black hole is among the black hole states, the most stable configuration, the most classical or semiclassical one. 

The Bekenstein-Hawking entropy $S_{sem}^{(0)}(J,Q)$ is minimal in the extremal case
\begin{equation}
S_{sem}^{(0)} (J, Q)_{extremal} = \frac{1}{2}~(1-\frac{\Delta^2}{2})~ S_{sem}^{(0)}(J=Q=0) 
\label{eq:Sextremal}
\end{equation}
with $\Delta $ given by Eq.~(\ref{eq:landa}). In particular, for the extremal Kerr black hole,
\begin{equation}
S_{sem}^{(0)} (J)_{extremal} = \frac{1}{2}~S_{sem}^{(0)}(J=0) 
\label{eq:SJextremal}
\end{equation}
A Kerr-Newman black hole cannot becomes through quantum decay an extremal black hole. The extremal black hole cannot be the late state of black hole evaporation. Through evaporation and decay, the black hole losses charge and angular momentum (super-radiance like processes)and at a higher rate than the loss of mass through thermal radiation. Thus, if a black hole was not extremal at its origin, it will not be extremal at its end. A Kerr Newman black hole becomes, through evaporation, a Schwarzshild black hole, at its late stage becoming a stringy state, which then decays, by the usual string decay process, in all types of massless and massive particles.
\section{Semi classical (Q.F.T) and quantum (string) regimes\label{sec:dual}}

From our previous analysis of the string canonical partition function Section ~(\ref{sec:partition}), the black hole quantum emission Section~(\ref{sec:qe}), and the string bounds on the black hole Section~(\ref{sec:bTJQ}), we see that a semiclassical black hole (BH) with size $L_{c\ell}$, mass $M$, temperature $ T_{sem}$, density of states $ \rho_{sem}$ and entropy $S_{sem}$, namely $(BH)_{sem}= (L_{c\ell}, M,  T_{sem}, \rho_{sem}, S_{sem})$, evolves through evaporation 
into a quantum string state of size $L_{s}$, mass $m$, temperature $T_{s}$, density of states  $\rho_{s}$ and entropy $S_{s}$, namely $(BH)_{s}$ = $(L_{s}, m , T_{s}, \rho_{s}, S_{s})$. 
The quantities in the set $(BH)_{sem}$ are precisely the semiclassical expressions of the respective ones in the set $(BH)_{s}$. In the quantum string regime, the black hole size $L_{c\ell}$ becomes the string size $(L_{s}$, the Hawking temperature $T_{sem}$ 
becomes the string temperature $T_s$, the black hole entropy $S_{sem}$ becomes the string entropy $S_s$.
The set $(BH)_{s}$ and $(BH)_{sem}$ are the same quantities but in different (quantum and semiclassical/cladssical) domains. That is, $(BH)_{s}$ is the quantum dual of the semiclassical set $(BH)_{sem}$ in the precise sense of the wave-particle (de Broglie)duality. This is the usual classical/quantum duality but in the gravity domain, which is {\it universal}, not linked to any symmetry or isommetry nor to the number or the kind of dimensions. 
From the semiclassical and quantum (string) black hole regimes $(BH)_{sem}$ and $(BH)_{s}$, we can write the semiclassical entropy $S_{sem}(M, J)$ for the Kerr black hole such that it becomes the string entropy $S_s(m,j)$ in the string regime, namely:

\begin{equation}
S_{sem}(M, J) = S_{sem}^{(0)}(M, J) - a ~k_B~\ln S_{sem}^{(0)}(J, M)~+~k_B~\ln~ F(S_{sem}^{(0)}, J)           
\label{eq:SsemBH}
\end{equation}

where $S_{sem}^{0}(M, J)$ is the Bekenstein-Hawking entropy given by Eq.~(\ref{eq:S0}) 
and $F(S_{sem}^{(0)}, J)$ is given by
\begin{equation}
F = \Delta^{-1} \Big(\frac{1 + \Delta}{2\Delta}\Big)^{a}~e^{\Big(\frac{1 -\Delta}{1 +\Delta}\Big) \frac{S_{sem}^{(0)}(M, J)}{\Delta k_B}} 
\frac{1}{\cosh^2 \Bigg(\frac{(1 - \Delta)}{\Delta}\frac{S_{sem}^{(0)}(M, J)}{2 k_B}\Bigg)} 
\label{eq:FBH}
\end{equation}

For $J=0$:~~~~~~$F = \bar{F} = 1$ ~~~~, ~~~~ $S_{sem}^{(0)}(M,J=0)\equiv S_{sem}^{(0)} = 4 \pi k_{B} \Big(\frac{M}{m_{Pl}}\Big)^2$
\\

$S_{sem}^{(0)}$ being the Schwarzschild black hole Bekenstein-Hawking entropy. $\Delta$ is given by Eq.~(\ref{eq:landa}), which in terms of $S_{sem}^{(0)}$, reads:
\begin{equation}
\Delta = \sqrt{1 - \Big(\frac{J}{\hbar}\Big)^2 \Big(\frac{4 \pi k_B~}{S_{sem}^{(0)}} \Big)^2}
\label{eq:deltaBH}
\end{equation}
Therefore, the whole Kerr entropy $S_{sem}(M, J)$ Eq.(~\ref{eq:SsemBH}) can be written enterely in terms of $S_{sem}^{(0)}$:
\begin{equation}
S_{sem}(M, J) = \Big(\frac{1 + \Delta^2}{2 \Delta} \Big) S_{sem}^{(0)}~ -~ a~k_B~\ln \Big (~\frac{S_{sem}^{(0)}}{k_B}~ \Big) - 
(a + 1)~k_B~ \ln\Delta ~-~ 2~k_B~ \ln~\cosh \Big [~  \frac{S_{sem}^{(0)}}{4 k_B}  \frac{(1 - \Delta^2)}{\Delta}~ \Big] 
\label{eq:SJFcos}
\end{equation}

The first term in Eq.~(\ref{eq:SJFcos}) reads 
\begin{equation}
\Big[~1 - \frac {1}{2}~\Big(\frac{J}{\hbar}\Big)^2 \Big(~\frac{4 \pi k_B}{S_{sem}^{(0)}}~\Big)^2~
\Big]~ {\Delta}^{-1}~ S_{sem}^{(0)}.
\label{eq:factorS}
\end{equation}
For $\Delta_s \neq0$, the effect of the angular momentum is to reduce the entropy. $S_{sem}(M, J))$
is maximal for $J=0$ (ie, for $\Delta = 1$). For $ J=0$, we have:
\begin{equation}
S_{sem}(M)~ = ~S_{sem}^{(0)}~-~ a ~k_B~\ln S_{sem}^{(0)}
\label{eq:Ssemzero}
\end{equation}
Notice the  {\bf new} term  $\ln F( S_{sem}^{(0)}(M,J))$  in Eq.~(\ref{eq:SsemBH}), enterely due to $J \neq 0$, and yielding in particular, to the last  $\ln cosh$  term in Eq.~(\ref{eq:SJFcos}). The argument of the last term in Eq.~(\ref{eq:SJFcos}) reads

\begin{equation}
X~\equiv~ \frac{S_{sem}^{(0)}}{4 k_B}\frac{(1 - \Delta^2)}{\Delta}~=~\frac{\pi}{\Delta}
\Big(\frac {J}{\hbar}\Big)^2\frac{4 \pi k_B}{S_s^{(0)}}~ =~ \frac{\pi}{\Delta}~ \Big(\frac{J}{\hbar}\Big)^2~ \Big(\frac{m_{Pl}}{M}\Big)^2
\label{eq:argX}
\end{equation}
\\
Eq.~(\ref{eq:SsemBH}) provides the whole Kerr black hole entropy $S_{sem}(M,J)$ as a function of the Bekenstein-Hawking entropy $S_{sem}^{(0)}(M,J)$. 
For $M\gg m_{Pl}$  and  $J < GM^2/c$,  $S_{sem}^{(0)}$  is the leading term of this expression, {\bf but} for high angular momentum, (nearly extremal or extremal case $J= GM^2/c$), a gravitational {\bf phase transition} operates and the whole entropy $S_{sem}$ is drastically  {\bf different} from the Bekenstein-Hawking entropy $S_{sem}^{(0)}$, as we precisely see in the Section below 

\section{A New feature : The Extremal Black Hole Phase Transition  \label{subsec:EXT}}

When $J$ reachs its {\bf maximal} value, that is $J = M^2 G/c^2$, $\Delta = 0$ and the term  $S_{sem}^{(0)}(M,J)$ is minimal :
\begin{equation}
S_{sem}^{(0)} (M,J)_{extremal} ~ = ~\frac{1}{2}~S_{sem}^{(0)}    
\label{eq:Sosemextrem}
\end{equation}
But for $\Delta \rightarrow 0$, the last term in Eq.~(\ref{eq:SJFcos}) substracts the first one, the pole in $\Delta$ of these two terms cancel out, and thus does  $S_{sem}^{(0)}$, yielding :
\begin{equation} 
S_{sem}(M,J)_{extremal} =   -(a + 1)~k_B~ \ln \frac {\Delta}{2}~ +~ k_B~\ln 2~ +~ \Delta ~\Big(~\frac{3} {4}~S_{sem}^{(0)}~-~ak_B~\Big)~+ ~ O(\Delta^{2})
\label{eq:Ssemextremal}
\end{equation}
In terms of the mass, or temperature:
\begin{equation} 
\Delta~ =~ \sqrt{1 - \Big(\frac{J}{\hbar}\Big)^2 \Big(\frac{m_{Pl}}{M} \Big)^4}~=~ \sqrt{1 - \Big(\frac{J}{\hbar}\Big)^2 \Big(\frac{T_{sem}}{T} \Big)^2} ~~,~~ T = \frac{1}{8\pi k_B}M c^2, 
\label{eq:DeltaM}
\end{equation}
$T_{sem}$ being the Schwarzshild Hawking temperature Eq.(~\ref {eq:TBH}).
In the extreme limit $(J/\hbar) \rightarrow (M/m_{Pl})^2 $ :
\begin{equation} 
\Delta_{extremal} = ~\sqrt{\frac {2}{M}}~\sqrt{M - \Big(\frac{J}{\hbar}\Big)^{1/2} {m_{Pl}}}~=~\sqrt{\frac {2}{T}}~\sqrt{T - \Big(\frac{J}{\hbar}\Big)^{1/2}T_{sem}}
\label{eq:DMextremal}
\end{equation}
and $S_{sem}(M,J)_{extremal}$ is dominated by
\begin{equation}
S_{sem}(M,J)_{extremal} =  -(a + 1)~k_B~ \ln ~(\sqrt{\frac {2}{T}}\sqrt{T - \Big(\frac{J}{\hbar}\Big)^{1/2} {T_{sem}}}~)~+~O(1)
\label{eq:SsemTextr}
\end{equation}
This shows that a {\bf phase transition} takes place at $T \rightarrow \sqrt{(J/\hbar)}~T_{sem} $, we call it {\bf extremal transition}. 
The characteristic features of this gravitational transition can be discussed on the lines of the extremal string transition we analysed for the extremal string states. Our discussion on the string case translates into the respective semiclassical black hole quantities  but we do not extend on these new features and implications here.

\begin{acknowledgements}

A. B. acknowledges the Observatoire de Paris, LERMA, for the kind hospitality extended to him.
M. R. M. acknowledges the Spanish Ministry of Education and Science (FPA04-2602 project) for financial support, and the Observatoire de Paris, LERMA, for the kind hospitality extended to her.

\end{acknowledgements}

\end{document}